\documentclass[aps,prd,10pt,notitlepage,nofootinbib,superscriptaddress,showkeys,showpacs]{revtex4-1}

\usepackage{amstext,amsmath,amssymb,amsfonts,bbm}
\usepackage[latin1]{inputenc}
\usepackage{epsfig}
\usepackage{hyperref}
\usepackage{amsthm}
\usepackage{subfigure}
\usepackage{color}
\usepackage{multirow}

\usepackage{graphicx}

\usepackage{latexsym}

\usepackage{cancel}

\newcommand{\bea}{\begin{eqnarray}}	
\newcommand{\eea}{\end{eqnarray}}
\newcommand{\beq}{\begin{equation}}	
\newcommand{\eeq}{\end{equation}}

\newcommand{\cG}{{\mathcal G}}
\newcommand{\cF}{{\mathcal F}}

\newcommand{\cL}{{\mathcal L}}

\newcommand{\kin}{\rm{kin\,}}
\newcommand{\inter}{{\rm int\,}}
\newcommand{\ext}{{\rm ext\,}}

\newcommand{\N}{{\mathbb N}}
\newcommand{\Z}{{\mathbb Z}}
\newcommand{\C}{{\mathbb C}}

\newcommand{\un}{{\rm uncolor}}
\newcommand{\col}{{\rm{color}}}

\newtheorem{lemma}{Lemma}

\newtheorem{theorem}{Theorem}
\newtheorem{proposition}{Proposition}

\begin{document}

\title{On the finite amplitudes for open graphs
\\ in Abelian dynamical colored Boulatov-Ooguri models}

\author{Joseph Ben Geloun}\email{jbengeloun@perimeterinstitute.ca}

\affiliation{Perimeter Institute for Theoretical Physics, 31 Caroline
St, Waterloo, ON, Canada \\
International Chair in Mathematical Physics and Applications, 
ICMPA-UNESCO Chair, 072BP50, Cotonou, Rep. of Benin}

\date{\small\today}

\begin{abstract}

In the work [Int.\ J.\ Theor.\ Phys.\  {\bf 50}, 2819 (2011)], it has been proved that the radiative corrections 
of the 2-point function in the $SU(2)$  Boulatov tensor model generates a relevant (in the Renormalization Group sense) contribution of the
form of a Laplacian. Such a term which was missing in the initial Boulatov model action should be 
added in that action before discussing the renormalization analysis of this model. 
In this work, by linearizing the group manifold, we prove that the amplitudes associated with Feynman graphs with external legs 
of the colored Boulatov model over $U(1)^3$ endowed with 
a Laplacian dynamics are all convergent. 
We conjecture that the same feature happens for the corresponding
Boulatov model over $SU(2)$. Higher rank models are also
discussed. 

\medskip
\noindent Pacs numbers: 11.10.Gh, 04.60.-m, 02.10.Ox
pi-qg-XXX and ICMPA/MPA/2013/009

\end{abstract}

\maketitle

\section{Introduction}

Tensor models \cite{ambj3dqg,mmgravity,sasa1,Ooguri:1991ib} are generalization of the famous matrix models \cite{Di Francesco:1993nw} 
which have a deep significance for quantum gravity in 2D. The Boulatov model \cite{Boul} is a particular tensor model 
which implements a sum over topologies of 3-manifolds supplemented by more singular objects. Such a model 
provides a second quantized formalism for quantum gravity in 3D \cite{Freidel:2005qe,oriti,Oriti:2006ar,Sasakura:2013gxg}. 
Recently, tensor models have been the focus of several investigations
due  to the acknowledged discovery 
of the 1/N expansion for a colored version of these tensor models \cite{Gur3,GurRiv,Gur4,Dartois:2013he,color,Gurau:2011xp,Gurau:2011kk,
Dartois:2013sra,Gurau:2013pca,gsch}. 
Based on this 1/N expansion,  the renormalization program for tensor models has been addressed
 in the framework of Tensorial Group Field Theories (TGFTs) \cite{Rivasseau:2011hm,Rivasseau:2012yp}. 
Successfully, it appears that several models prove to be renormalizable
\cite{BenGeloun:2011rc,Geloun:2012fq,BenGeloun:2012pu,Geloun:2012bz,Carrozza:2012uv,
Carrozza:2012uv,Samary:2012bw,Carrozza:2013wda,Geloun:2013saa} and even genuinely asymptotically free
 \cite{BenGeloun:2012pu,BenGeloun:2012yk}\cite{Samary:2013xla,Geloun:2013saa}.

The above mentioned tensor field models have particular kinetic terms.  
Some of these models possess just a mass term while others have higher order derivatives encoding their dynamics.  
A specific choice of a kinetic term turns out to be crucial for the renormalization analysis. 
Indeed, the fact that a Laplacian in the kinetic term was important for carrying a renormalization procedure for the 
Boulatov-Ooguri tensor models has been highlighted in \cite{Geloun:2011cy}.
In the following, we will focus and review this particular 
contribution simply because the present work must
be regarded as a sequel of it. 

The Boulatov model is defined with integrable functions $\phi(g_1,g_2,g_3)$ as fields living on three copies of the group $G=SU(2)$, $g_i \in SU(2)$. Using the spin $j$
momentum representation space of $SU(2)$, one realizes that a field $\phi$ is associated with a rank 3 tensor. 
Especially, the functions $\phi$ satisfy a particular invariance
$\phi(g_1h,g_2h,g_3h)=\phi(g_1,g_2,g_3)$, $\forall h \in SU(2)$,
which makes their associated Fourier tensor components as $SU(2)$ invariants. The important features of tensor models 
lie in the dual counterparts of their ingredients. 
For the above model, the fields $\phi$ are dual to 2-simplexes 
(or triangles) and the interaction is defined via the 
product of four fields with arguments paired in a special way 
representing a 3-simplex (or a tetrahedron). 
For arbitrary rank, these ideas generalize without ambiguity.
At the perturbative level, the Boulatov model partition function expands in terms of graphs or 
equivalently simplicial complexes in 3D where tetrahedra are glued along their boundary triangles. The graph amplitudes in 
this setting may diverge and, as such, must be regularized either
by heat kernels or by imposing a sharp cut-off $\Lambda$ on all 
large spins. Starting with the Boulatov model with simply a mass term $m^2\phi^2$ and interaction $\lambda \phi^4$, 
the authors of \cite{Geloun:2011cy} expanded in power series of the cut-off $\Lambda$ in the spin momentum space, 
the initial divergent amplitude of the 2-point function made with $V$ vertices.
As a result, it turns out that there exist two divergent corrections: 
a 0th order correction which should be considered as a mass renormalization but also, quite surprisingly, 
a second order correction of the rough form $\Lambda\sum_{s=1}^3\phi \Delta_s \phi$, where $\Delta_s$ 
is a Laplace operator on the group manifold acting on one strand of the tensor field $\phi(g_1,g_2,g_3)$. Consequently, the presence of such a correction suggests that we should
introduce in the kinetic term a Laplacian term such that this diverging ``orphan''  
correction would be associated to a wave function renormalization of the model.

The purpose of this work stems exactly from that idea to investigate the effect of introducing a Laplacian in the 
Boulatov model action from the very beginning. We refer this class of models to as ``dynamical''
in distinction with models with just a trivial mass term in the kinetic
term. In this work, we investigate an Abelian version of the Euclidean Boulatov-Ooguri models that is we replace, in these class of models, the group $SU(2)$ by $U(1)^{3}$.\footnote{Such a prescription was called ``group linearization'' in \cite{Geloun:2010nw} and turned out to be useful for understanding the divergence of the graph amplitudes in
terms of the homology associated with the simplicial structure of the graphs.} Inserting a Laplacian in the kinetic term and using the 
Gurau colored prescription on tensor fields \cite{color} which allows us to have a control
of the combinatorics of Feynman graphs, we prove that all amplitudes for open graphs,  i.e. graphs with external legs (simplicial complexes with boundary), of the rank 3 (Boulatov) model are convergent. 
Since the power counting theorems for tensor models over an Abelian group and a non-Abelian group with the same dimension are very similar \cite{Carrozza:2013wda}, we conjecture that the above statement will be also true for the  dynamical colored Boulatov model over $SU(2)$. 
As a consequence of the present study, one may finally conclude that:

-  the (Abelian) colored Boulatov model without Laplacian dynamics is
perturbatively non-renormalizable;

-  the (Abelian) colored Boulatov model with Laplacian dynamics have
 finite open amplitudes.

Let us quickly discuss the reason why open graph amplitudes are important. 
Renormalization is about understanding the amplitude divergences with respect to some scale \cite{Rivasseau:1991ub}. In the standard
quantum field theory renormalization procedure, the type of diverging
 contributions that should be subtracted are those who look local from the point of their external legs. In other words, given a locality principle
(attached to a set of initial terms or invariant data), only graphs with internal scale much higher with respect to some external scale and which satisfy the 
locality principle should contribute to the coupling renormalizations. Thus, the class
of graphs which are important in this context is definitely the class of 
open graphs for which external scales  
can be distinguished from internal ones. 

The plan of the present work is as follows. We define, in the next section,
the rank 3 model under consideration and prove our main result,
namely, Theorem \ref{theo:ampl}. Section \ref{sect:rnkd}
addresses the general rank $d$ situation. The last section gives a summary of our results and further discusses our results.

\section{The dynamical Abelian colored Boulatov model}
\label{model}

\noindent{\bf The model -} The fields $\phi^a$, $a=0,1,2,3$, belong to the Hilbert space of square integrable complex functions on $U(1)^{3}$ which  satisfy the gauge invariance condition
\begin{equation}
\phi^a(g_1h,g_2h,g_3h)=\phi^a(g_1,g_2,g_3)\;, \quad \forall g_i, h \in  
U(1)^3\;. \label{inv}
\end{equation}
We will use henceforth shorthand notation $\phi^a(g_1,g_2,g_3) = \phi^a_{1,2,3}$ and parametrize as usual $g_s =e^{i\vec \theta_s}\in U(1)^3$,
with $\vec \theta_s = (\theta_{s,1},\theta_{s,2},\theta_{s,3}) \in [0,2\pi)^3$. The index $a$ is referred to as the color of the field. 
The field is not assumed to be symmetric under permutation of its 
arguments. 

The action is formed by a kinetic term and interaction part. The kinetic term is given by
\begin{equation}
S^{\kin}:= \int [\prod_{\ell=1}^{3}dg_\ell]\ \left[ 
\sum_{a=0}^3\bar\phi^a_{3,2,1}\Big(-\sum_{s=1}^3\Delta_{s}
+m^2 \Big) \phi^a_{1,2,3}\right],
\label{kine}
\end{equation}
where $dg_s=d\vec\theta_s/(2\pi)^3$ denotes the Haar measure on $U(1)^3$, the operator
$\Delta_{s}= \partial_{\vec\theta_s}\partial_{\vec\theta_s}
= \sum_{p=1}^3 \partial_{\theta_{s,p}}\partial_{\theta_{s,p}}$ represents the Laplacian which acts on the strand argument indexed by $s$.

Dually associated with a $3$-simplex, the interaction in the colored Boulatov model is nonlocal and reads
\begin{equation}
S^{\inter}[\phi,\bar\phi]:=\lambda\int [\prod_{\ell=1}^{6}dg_\ell]\Big( \phi^0_{1,2,3}\,
\phi^1_{3,4,5}\,\phi^2_{5,2,6}\,\,\phi^3_{6,4,1}
 + \bar\lambda \ \bar\phi^0_{1,2,3}\,
\bar\phi^1_{3,4,5}\,\bar\phi^2_{5,2,6}\,\,\bar\phi^3_{6,4,1}\Big)\;,
\label{intera}
\end{equation}
with a particular pairing of the six variables 
according to the pattern of the edges of a tetrahedron.

At the quantum field level, we are interested in the partition
function
\beq
Z = \int d\mu_C(\phi,\bar\phi) \, e^{-S^{\inter}[\phi,\bar\phi]}\,,
\label{zcolo}
\eeq
where $d\mu_C(\phi)$ is a Gaussian measure with covariance
dictated by the kinetic term \eqref{kine} and the gauge invariance
on fields \eqref{inv} such that 
\bea
C_{ab}(\{g_i\};\{\tilde g_i\}) &=&  \int d\mu_C(\phi,\bar\phi) \;
\bar\phi^a(g_1,g_2,g_3)\phi^b(\tilde g_1,\tilde g_2, \tilde g_3)
 \crcr
&= & 
\sum_{\vec q_s \in\Z^3}\int_{[0,2\pi)^3} d\vec\lambda \int_0^{\infty} d\alpha\;
e^{-\alpha(\sum_{s=0}^3 |\vec q_s|^2 + m^2)}
e^{\iota\sum_{s=0}^3 \vec q_s \cdot [\vec\theta_s - \vec\lambda + \vec{\tilde{\theta}}_s]} \; \delta_{a,b}\,,
\label{cab}
\eea
where we introduce the momentum variable $\vec q_s \in \Z^3$ for each 
argument $g_s$ of the fields. More details on \eqref{cab} can be found in \cite{Carrozza:2012uv}.  
Note that only are allowed contractions between conjugate tensors
with the same color. 

\begin{figure}[h]
 \centering
     \begin{minipage}{.5\textwidth}
\includegraphics[angle=0, width=8cm, height=1cm]{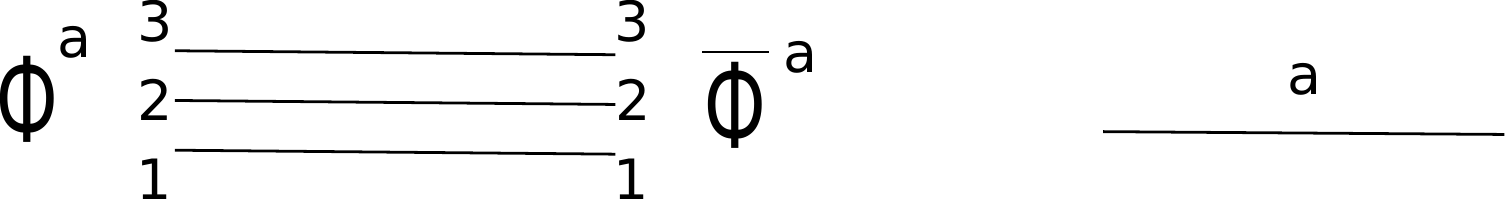} \\
\vspace{0.4cm}
\includegraphics[angle=0, width=8cm, height=2.2cm]{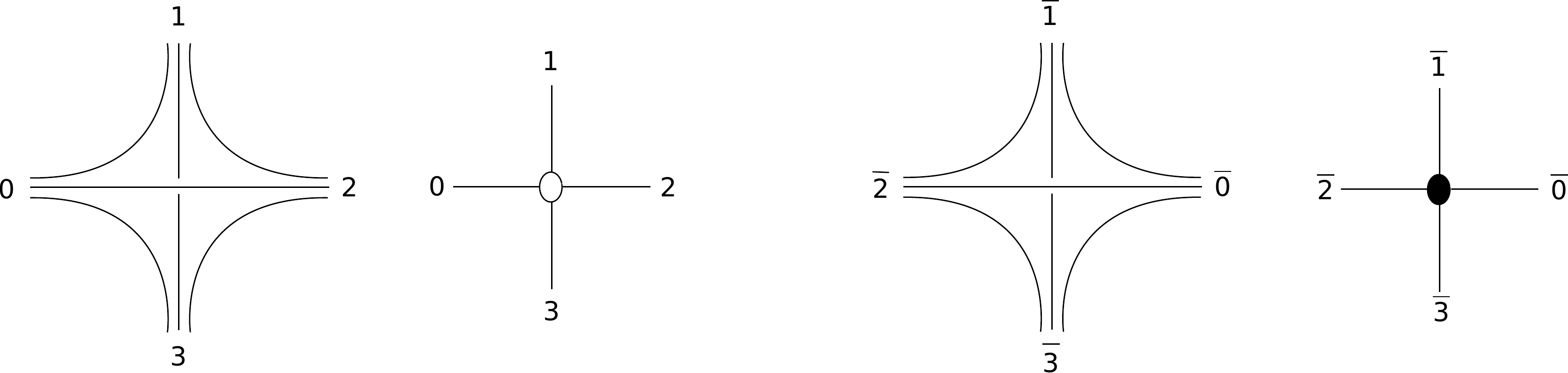} 
\caption{ {\small Propagator (top) and vertices (bottom) of the model
and their different graphical representations. }} \label{fig:v4d}
\end{minipage}
\end{figure}

Feynman graphs in this theory are obtained from 
the gluing of two types of vertices (say black and white) by propagator lines (see Figure \ref{fig:v4d}).
The resulting graph is stranded and possesses also a line coloring induced by the colors of the fields contracted. 
The graph is also bipartite because of the  presence of ``conjugate'' vertices and this induces an orientation 
on each line (for instance exiting from the black vertices and entering the white ones). 
There is therefore a colored and bipartite graphical
way to represent all Feynman graphs in this theory. 
A face in the graph is a single strand connected component. 
A face is called closed according to the fact that
the strand which defines it closes as a loop, otherwise it is called open.  
Since the model is colored, to each face one can associate
a pair of colors $(a,b)$ \cite{color}.  
One can also give an orientation to a single strand or face 
by simply putting an arbitrary arrow on it. Using these face and line orientations, 
the notion of incidence matrix of a stranded graph can be consistently defined by the data 
$\{\epsilon_{lf}\}_{l \in \cL; f \in \cF}$, where $\cL$ is the set of lines of the graph and 
$\cF$ is the set of faces of the graph, such that 

- $\epsilon_{lf}=+1$, if the face $f$ passes through the line $l$ (we write $l \in f$) and their orientation coincide

- $\epsilon_{lf}=-1$, if $l \in f$ and their orientation do not match 

- $\epsilon_{lf}=0$ if the face $f$ does not pass through the line $l$
(we write $l \notin f$). 

The rank $R$ of the matrix $\epsilon_{lf}$ is an important ingredient for the following. We write $|\cL|=L$ and $|\cF|=F$.

\medskip 

\noindent{\bf Slice decompostion -} 
The next stage is to introduce a slice decomposition of the propagator kernel such that
\bea
C_{ab} = \sum_{i=0}^\infty C_{ab}^i \,, 
\qquad 
C_{ab}^i = \sum_{\vec q_s }\int d\vec\lambda \int_{M^{-2i}}^{M^{-2(i-1)}} d\alpha\;
e^{-\alpha(\sum_{s=0}^3 |\vec q_s|^2 + m^2)}
e^{\iota\sum_{s=0}^3 \vec q_s\cdot [\vec\theta_s - \vec\lambda + \vec{\tilde{\theta}}_s]} \; \delta_{a,b}\,,
\eea
where $M>0$. $C^\rho_{ab}=\sum_{i\leq \rho}C^i_{ab}$ refers 
to the regularized UV propagator where $\rho$ is the cut-off 
and $C^0_{ab}$ to the IR propagator slice. 

The following bound holds \cite{Carrozza:2012uv}: $\forall i \in \N$,
\beq
|C^i_{ab}(\{\vec\theta_s\};\{\vec{\tilde{\theta}}_s\})| \leq 
K M^{7i}\int d\vec\lambda\;
e^{ -\delta\, M^{-i} \sum_{s=0}^3 |\vec\theta_s - \vec\lambda + \vec{\tilde{\theta}}_s|} \; \delta_{a,b}\,.
\label{bound0}
\eeq
for some constants $K$, $\delta$.  

\medskip 

\noindent{\bf  Divergence degree -}
At this stage, all renormalizable tensor models possess interaction terms  built from the so-called 
melonic interactions \cite{Bonzom:2011zz} with uncolored fields \cite{Bonzom:2012hw}. 
The resulting graphs in uncolored models can be thought as graphs obtained after integration of all but one color in the above
partition function $Z$ and then choosing that last color (say 0) to be the 
one possessing a non trivial dynamics. The power counting theorem in such models derives 
from a multi-scale analysis and leads to a divergence degree
for any {\it face--connected} graph $\cG$ (namely a graph for which 
the matrix $\epsilon_{lf}$ does not factorize by block), as 
\cite{Carrozza:2012uv}:
\beq
\omega_d^{\un}(\cG) = -2 L + \dim G (F- R)\,,
\label{omun}
\eeq
where $\dim G$ is the dimension of the group manifold.

In fact, after a direct inspection of the multi-scale analysis leading to this statement in \cite{Carrozza:2012uv}, 
the result essentially depends on the form of the graph amplitude which factorizes in terms of external and internal faces. 
The same happens for the class of models  described by \eqref{kine}
and \eqref{intera} or their extension for any rank $d$. Finally, performing step by step a 
similar multi-scale analysis for the present model using \eqref{bound0} leads to a divergence 
degree $\omega_d^{\col}(\cG)$ in a exact similar expression as given by \eqref{omun}. In short, 
the multi-scale analysis does not depend on the fact that the model is uncolored. 
\begin{proposition} The superficial degree of divergence of a face-connected graph $\cG$ in the colored model is given by 
\beq
\omega_d^{\col}(\cG) = -2 L + \dim G (F- R)\,,
\label{omun2}
\eeq
where $\dim G$ is the dimension of the group manifold
and $d$ the theory rank. 
\end{proposition}

The main result of this paper is given by the following statement: 

\begin{theorem}[Finite open graph amplitudes]\label{theo:ampl}
For any face-connected graph $\cG$ with number $N_{\ext}\geq 2$ of external 
legs  in the model described by \eqref{kine} and \eqref{intera}, we have 
\beq
\omega_3^{\col}(\cG) <0\,. 
\eeq
\end{theorem}

\medskip 

In order to prove this theorem, we must introduce
two concepts: 

(1) Faces passing by $k$ lines (called $k$-faces) which do not need 
further comments; we denote $\cF_{\inter; k}$ the set of $k$-faces
which are closed and $F_{\inter;k}=|\cF_{\inter; k}|$. Since, we have a bipartite theory, 
the number of edges visited by the closed faces must be even and so, here, $k$ is an even integer strictly 
larger than 0. We denote also $\cF_{\ext; k}$ the set of $k$-faces which are open and $F_{\ext;k}=|\cF_{\ext; k}|$.
In the latter situation, $k$ can be an arbitrary positive integer even 0. 
The total number of internal and external faces can be written 
\beq\label{faces}
F_{\inter} = \sum_{k\geq 2/ k \in 2\N} F_{\inter; k}\,,
\qquad 
F_{\ext} = \sum_{k\geq 0} F_{\ext; k}\;. 
\eeq

\begin{figure}[h]
 \centering
     \begin{minipage}{.7\textwidth}
\includegraphics[angle=0, width=6cm, height=1cm]{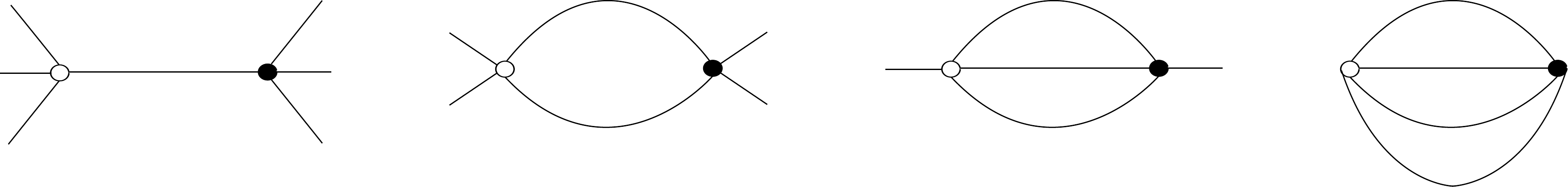} 
\caption{ {\small 1,2,3,4-dipoles, respectively. }} 
\label{fig:dip}
\end{minipage}
\end{figure}

(2) A $p$-dipole of some graph $\cG$, with $p=1,2,3,4$, 
 can be simply thought as a subgraph of $\cG$ or a subset of $p$ lines
joining a pair of vertices (see Figure \ref{fig:dip}). A 4-dipole must be a disconnected component graph. Since we are only interested in
graph with external legs, we will not take into account 
these 4-dipole contributions. We denote $D_p$ the number
of $p$-dipoles in $\cG$. 

\medskip

We shall establish two lemmas.

\begin{figure}[h]
 \centering
     \begin{minipage}{.7\textwidth}
\includegraphics[angle=0, width=5cm, height=2.5cm]{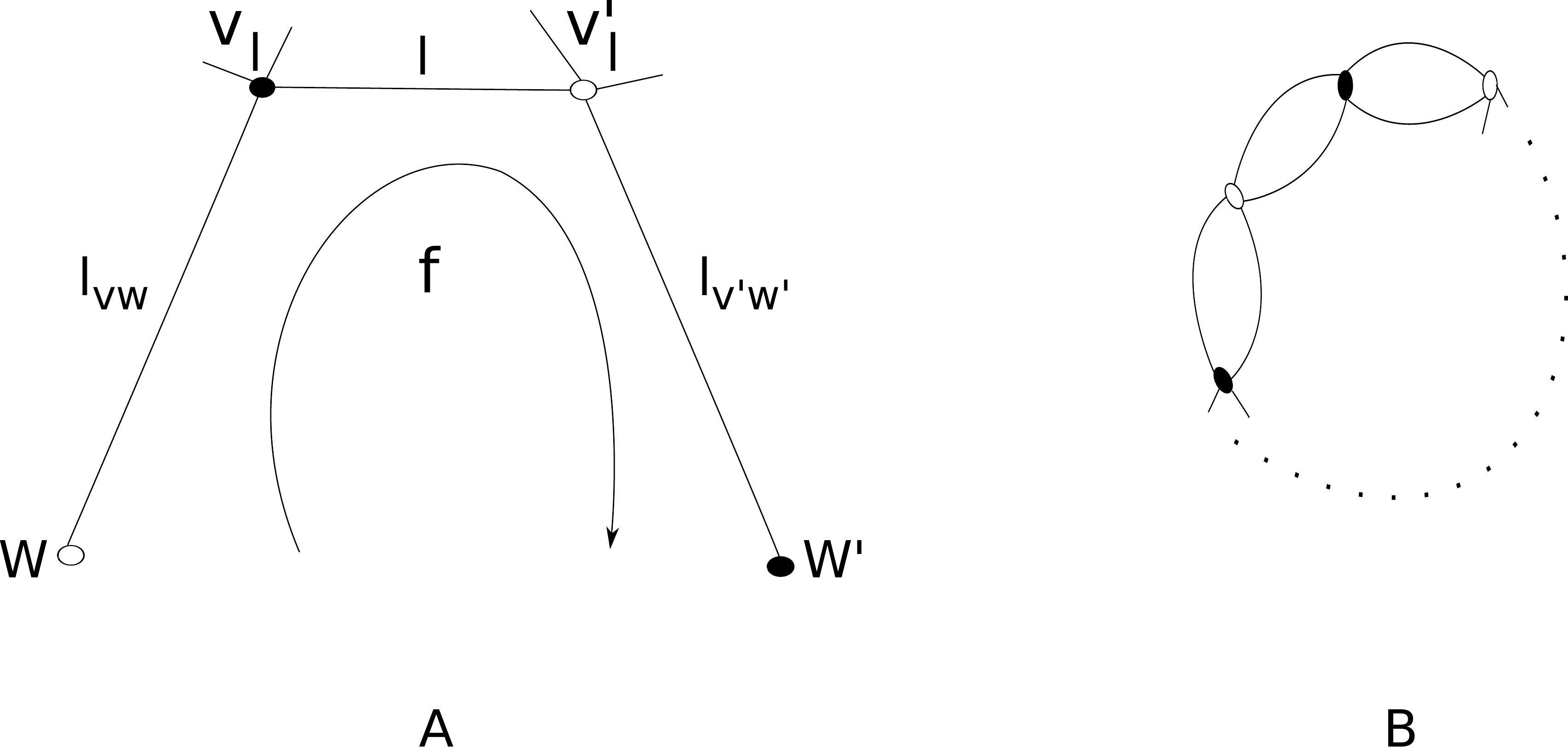} 
\caption{ {\small A line $l$ in a $k$-face (A) and the closed graph formed with 2-dipoles (B). }} 
\label{fig:kface}
\end{minipage}
\end{figure}

\begin{lemma}[$k$-face]\label{lem:4f}
Consider a graph $\cG$, an integer $k\geq 4$ and 
a closed $k$-face $f$ in $\cG$. Either there is at least one line of 
$f$ which is not a $p$-dipole line of $\cG$, $p=2,3$, 
or $f$ belongs to a disconnected component vacuum subgraph
of $\cG$.  
\end{lemma}
\proof  Consider a closed $k$-face $f$ with $k \geq 4$ such that 
all lines where $f$ passes are $p$-dipole lines. 
Pick any line $l\in f$ and consider its incident vertices
$v_l$ and $v'_l$. Since $k\geq 4$ and that the graph is bipartite, there must be at least two  other vertices, $w$ and $w'$ linked to $v_l$ and $v'_l$,
by some lines $l_{vw}\in f$ and $l_{v'w'}\in f$, respectively, such that we have a configuration given by Figure \ref{fig:kface} A. 
Consider other half-lines exiting from $v_l$.  
Since we assume that $l$ is a dipole line, all these half-lines should
be fully contracted  either with $v'_l$ or with $w$. 
Note that if $l$ belongs to a 3-dipole, it is immediate
that $l_{vw}$ is not a 2,3-dipole. That contradicts our hypothesis
that all lines where $f$ passes must be dipoles. 
Thus, between $v_{l}$ and $v_{l}'$, there must be at most
a 2-dipole. This holds for any line $l$ of the graph where $f$ passes. 
We have a unique possible configuration for connecting
the $k$ vertices of the lines of $f$ and the rest of the colored lines for making a bipartite and colored 
graph so that all lines of $f$ are included in a 2-dipole, it is given by Figure \ref{fig:kface} B. 
Thus, $f$ belongs to this component which is a disconnected vacuum graph.  

\qed

\begin{lemma}[Lower bound on $R$]\label{lem:br}
For a face-connected open graph $\cG$, the rank $R$ of the $\epsilon$ matrix is bounded from below as
\bea
R \geq D_2 + 2 D_3 + \frac13 F_{\inter; \geq 4} \,, 
\eea
where $F_{\inter; \geq q}:=\sum_{k\geq q}  F_{\inter; k}$. 
\end{lemma}
\proof The strategy that we use is somehow similar as the one developed in \cite{Carrozza:2012uv}. 
Our goal is to extract the maximum rank from the $\epsilon$ matrix using the structure of $\cF_{\inter; k}$. 
Consider an open face-connected graph $\cG$. Each $p=2,3$-dipole of $\cG$  generates a certain number of 
closed faces -- 1,2 for $p=2,3$, respectively -- which do not pass through any other lines of the graph.

 Consider the matrix $\epsilon$ where we arrange the lines 
and columns in the following way: we first list the faces contributing
to $F_{\inter; 2}$, then $F_{\inter; 4}$, etc, in rows. 
Noting now that the 2-faces elements of $\cF_{\inter; 2}$ appear only in $p$-dipoles, $p=2,3$, 
we arrange the columns by listing the (2,3)-dipole lines $\ell^{(2)}=\{\ell^{(2)}_a\}_a$ where these 2-faces pass. 
Then, we insert the lines $\ell^{(4)}=\{\ell^{(4)}_b\}_b$ where the 4-faces belonging to $\cF_{\inter; 4}$ pass, etc. 
One must obtain the following matrix structure:
\beq
\begin{array}{c|c|c}
 &\ell^{(2)} & \ell^{(>2)}  \\ \hline
\cF_{\inter; 2} & \epsilon^{(2)} &  0  \\ \hline
\cF_{\inter; 4} &  \ast \, \cdots \, \ast \,  & \epsilon^{(4)}  \\ \hline
\cF_{\inter; > 4} &\ast \, \cdots \, \ast \,  &\epsilon^{(>4)} \\
\end{array}
\label{tab}
\eeq

Focusing on the block $\epsilon^{(2)}$ of the $\epsilon$ matrix, it is direct to realize that, 
from the faces contributing to $\cF_{\inter; 2}$, a minor of rank   
\beq
D_2+ 2 D_3 
\eeq
 can be extracted. We can now erase all columns contributing to $D_{2,3}$ or equivalently $\ell^{(2)}$, from the matrix. 

Let us focus on the rest $F_{\inter; \geq 4}$. From Lemma \ref{lem:4f},
after erasing the lines coming from $\ell^{(2)}$, the rows labeled
by $\cF_{\inter;\ge 4}$ cannot contain only 0's, otherwise that would
mean that all elements of $\cF_{\inter;4}$ were contained
in 2-dipole lines making a disconnected component. Another observation is that any line having three strands
(indexed by a pair of colors $(a,b)$) defines necessarily 3 different faces. 
So, in order to extract the maximum of rank in $\epsilon^{(\geq 4)}$,
we proceed as follows: we choose a face $f$ in $\cF_{\inter;\geq 4}$
and one of the remaining line $\ell_f \in \ell^{(>2)}$ where this face passes through. 
We erase the rows defined by the other closed faces which could 
pass by  $\ell_f$. If there are still some faces remaining in $\cF_{\inter; \geq 4}$, we continue the algorithm.  
At the end, one can extract a minor in $\epsilon^{(\geq 4)}$ of rank at least  bounded by $F_{\inter; \geq 4}/3$.  

\qed 

We are in position to prove Theorem \ref{theo:ampl}. 

\proof[Proof of Theorem \ref{theo:ampl}] 
Each internal line produces three strands used either in internal faces or external ones. Thus, 
\beq\label{lines}
3 L = \sum_{k \geq 2/ k \in 2\N} k F_{\inter,k} + \sum_{k\geq 1} k F_{\ext,k} \,. 
\eeq
Keeping in mind that the number of faces of a connected graph $\cG$
decomposes in terms of faces passing through  exactly $k$ lines
as given in \eqref{faces}, we return to the degree of divergence and insert \eqref{faces} and \eqref{lines} in \eqref{omun2} and obtain:
\beq
\omega_3^{\col}(\cG)
= \sum_{k \geq 2/ k \in 2\N} \Big( 3- \frac{2k}{3}\Big) F_{\inter,k} 
-\sum_{k\geq 1} \frac{2k}{3} F_{\ext,k} - 3R \,.
\eeq
One recognizes that the possible sources of divergences come from 
the presence of $F_{\inter,k}$, for $k=2,4$. 
We further decompose  $F_{\inter;2}$ 
in terms of the numbers $D_p$ of $p$-dipoles, $2 \leq p \leq 3$, as
\beq
F_{\inter, 2} =  D_2 + 3D_3\,,
\eeq
and use Lemma \ref{lem:br} to re-express and bound
the divergence degree as 
\bea\label{omeco}
\omega_3^{\col}(\cG)
& \leq &
\sum_{k \geq 2/ k \in 2\N} \Big( 3- \frac{2k}{3}\Big) F_{\inter; k} 
-\sum_{k\geq 1} \frac{2k}{3} F_{\ext; k} - 3
(D_2+ 2 D_3 + \frac13 F_{\inter,4} +\frac13 F_{\inter; >4}  ) 
\crcr
&\leq & 
- \sum_{k \geq 6/ k \in 2\N} \Big(  \frac{2k}{3}-3\Big) F_{\inter,k} 
-\sum_{k\geq 1} \frac{2k}{3} F_{\ext,k} - 
(\frac{4}{3} D_2+  D_3 ) -\frac23  F_{\inter,4} \,.
\eea
Since the graph is assumed to be open then 
$\sum_{k\geq 1} \frac{2k}{3} F_{\ext,k}>0$ and this
achieves the proof of the theorem. 

\qed

Hence, open graph amplitudes in this dynamical colored model
are all finite. There is no need to renormalize the amplitudes in this
setting. Other implications of this fact will follow in the discussion in
the  conclusion. 

\section{Rank $d$ colored Boulatov-Ooguri models}
\label{sect:rnkd}

It is instructive to scrutiny how  the previous analysis
extends to any tensor models over Abelian groups built over $U(1)$ and to  arbitrary theory rank. In order to proceed, we 
consider a group $G=U(1)^D$ and the Boulatov-Ooguri 
models of rank $d$ which are defined via colored fields $\varphi^{a}:G^d \to \C$,
$a=0,1,\dots, d,$ satisfying the generalized version of \eqref{inv}. 
The action which generalizes \eqref{kine} and \eqref{intera} can be inferred in a direct manner. The pattern of the interaction follows
a $d$-simplex now (see \cite{Geloun:2009pe} for graphical representations). 

At the quantum level, the path integral keeps its form given by 
\eqref{zcolo} with the field measure $d\mu_C$ now defined by 
the covariance
\bea
C_{ab}(\{g_i\};\{\tilde g_i\}) &=&  \int d\mu_C(\phi,\bar\phi) \;
\bar\phi^a(g_1,\dots ,g_d)\phi^b(\tilde g_1,\dots , \tilde g_d)
 \crcr
&= & 
\sum_{\vec q_s \in\Z^3}\int_{[0,2\pi)^D} d\vec\lambda \int_0^{\infty} d\alpha\;
e^{-\alpha(\sum_{s=0}^d |\vec q_s|^2 + m^2)}
e^{\iota\sum_{s=0}^d \vec q_s \cdot [\vec\theta_s - \vec\lambda + \vec{\tilde{\theta}}_s]} \; \delta_{a,b}\,.
\eea
In a slice $i$, the propagator can be bounded as \cite{Carrozza:2012uv} (see Proposition 5 therein)
\beq
|C^i_{ab}(\{\vec\theta_s\};\{\vec{\tilde{\theta}}_s\})| \leq 
K M^{(dD-2)i}\int d\vec\lambda\;
e^{ -\delta\, M^{-i} \sum_{s=0}^d |\vec\theta_s - \vec\lambda + \vec{\tilde{\theta}}_s|} \; \delta_{a,b}\,,
\label{bound3}
\eeq
for some constants $K$, $\delta$. As discussed in the previous section, the general multi-scale analysis and power counting theorem yield the degree of divergence
of a graph $\cG$ as 
\beq
\omega_d^{\col}(\cG) = -2 L + \dim G (F- R)\,. 
\label{omun3}
\eeq
The main basic lemmas \ref{lem:4f} and \ref{lem:br} find an extension
in the present setting as follows:

\begin{lemma}[$k$-face II]\label{lem:4fd} Consider a graph $\cG$, an integer $k\geq 4$ and a closed $k$-face $f$ in $\cG$.
Either there is at least one line of $f$ which is not a $p$-dipole line of $\cG$, $p=2,3, \dots, d$, 
or $f$ belongs to a disconnected component vacuum subgraph of $\cG$.  
\end{lemma}

\begin{lemma}[Lower bound on $R$ II]\label{lem:brd}
For a face-connected open graph $\cG$, the rank $R$ of the $\epsilon$ matrix is bounded from below as
\beq
R \geq \sum_{k=2}^{d}  (k-1)D_k + \frac1d F_{\inter; \geq 4} \,, 
\eeq
where $F_{\inter; \geq q}:=\sum_{k\geq q}  F_{\inter; k}$. 
\end{lemma}

\proof[Proof of Lemma \ref{lem:4fd}]
The proof of Lemma \ref{lem:4fd} follows from adjusting the proof
of Lemma \ref{lem:4f}. First, $(d+1)$-dipoles are vacuum graphs
and should be excluded from the discussion. Consider now 
that all lines of $f$ are contained in dipoles. Consider any
such line $l\in f$, 
$v_l$ and $v'_l$, $w$ and $w'$ vertices and lines $l_{wv}$ and $l_{v'w'}$ as defined in Lemma \ref{lem:4f}'s proof. $f$ cannot belong to a  $d$-dipole
otherwise $l_{vw}$ and $l_{v'w'}$ are not $2,\dots, d$-dipole lines.  
Then, using again the fact that between $v_l$ and $v_l'$ there should be
at most a $(d-2)$-dipole and, so, some their half-lines are hooked to $w$ and $w'$, one arrives again at a construction of a vacuum graph
containing $f$ if all lines passing through $f$ are supposed to be in dipoles. 
\qed

\proof[Proof of Lemma \ref{lem:brd}] We proceed in the same 
way as in Lemma \ref{lem:br}. After arranging the $\epsilon$ 
matrix as given in \eqref{tab}, we first extract a minor of rank
\beq
\sum_{k=2}^{d}  (k-1)D_k
\eeq
due to the presence of $2, \dots, d$-dipoles.  
We can now erase the columns $\ell^{(2)}$. 
Next, we extract a minor from the remainder faces 
$F_{\inter; k\geq 4}$ which cannot be totally vanishing because
of Lemma \ref{lem:4fd}. Using the fact that lines pass exactly 
through $d$  different faces, we use the same recipe as before
and collect one face per line and erase the remaining rows. 
At the end, in $\epsilon^{(\geq 4)}$, we can gain a minor of rank bounded by 
\beq
\frac{1}{d} F_{\inter; \geq 4}\,. 
\eeq

\qed

The goal is to obtain a bound on the divergence degree
in this general instance. In order to proceed with this, 
we decomposes the lines in strands and obtain the relation: 
\beq\label{dL}
d L = \sum_{k \geq 2/ k \in 2\N} k F_{\inter,k} + \sum_{k\geq 1} k F_{\ext,k} \,.
\eeq
The number of internal 2-faces expanded in terms of dipoles will be also needed:  
\beq\label{f2d}
F_{\inter; 2} = \sum_{k=2}^{d} \frac{k(k-1)}{2}D_k \,.
\eeq
Using \eqref{dL}, \eqref{f2d} and Lemma \ref{lem:brd}, 
we re-evaluate the divergence degree \eqref{omun3} 
and bound it as:
\bea
\omega_d^{\col}(\cG)
& =& -\frac{2}{d}\Big( \sum_{k \geq 2/ k \in 2\N} k F_{\inter,k} + \sum_{k\geq 1} k F_{\ext,k}\Big) + D \sum_{k \geq 2/ k \in 2\N}  F_{\inter,k} - DR \crcr
&\leq & 
-\sum_{k \geq \frac{Dd}{2}/ k \in 2\N} \Big( \frac{2k}{d}-D\Big) F_{\inter,k} 
-\sum_{k\geq 1} \frac{2k}{d} F_{\ext,k}\crcr
&+& \sum_{k=2}^{d}\Big\{ \Big( D- \frac{4}{d}\Big) \frac{k}{2} 
- D \Big\} (k-1)D_k 
+
\sum_{4\leq k <  \frac{Dd}{2}/ k \in 2\N}   \frac{Dd-2k-D}{d} F_{\inter; k} 
\label{bound}
\eea
A rapid checking proves that, setting $D=3$ and $d=3$ in 
\eqref{bound}, one recovers \eqref{omeco}.  
Following our previous practice, we investigate if, term 
by term, the above quantity is strictly negative. The two first sums are 
obviously negative. Next, under the condition 
\beq
Dd > 8\,, \qquad   8 \leq  2k  <  Dd
\eeq
we have 
\beq
- D < Dd-2k-D \leq  D(d-1)-8
\eeq
which can lead to positive $Dd-2k-D$ for some enough large
value of $D$ and $d$. On the other hand,
for $ 2 \leq k \leq d$
\beq
-2 \leq  \Big( D- \frac{4}{d}\Big) \frac{k}{2}  -D\leq \frac{Dd^2 -8-2dD}{2d} 
\eeq 
which can be negative or positive. In conclusion, the amplitudes
in any rank $d$ model can be both convergent and divergent. 

Note by constraining the upper
bounds of these inequalities, namely by requiring 
\beq
D \geq 1 \,, \quad d \geq 3 \,, \quad 
D(d-1)\leq 8 \quad 
\text{ and } \quad  Dd(d-2) \leq 8
\eeq 
we can quickly reach several 
colored models over $U(1)^{D}$ which have finite open amplitudes.
For instance, the Abelian rank $d$ colored models given by 
\beq
(\,d= 3 ,\, G = U(1) \,),  \qquad  
(\,d= 3 ,\,  G = U(1)^{2} \,), \qquad \text{ and } \qquad 
(\, d= 4 , \,  G = U(1)\,) \,
\eeq
have all convergent amplitudes. In fact, the requirement that the quantity 
\eqref{bound} is negative term by term is clearly too strong
for having an overall convergent amplitude. 
Furthermore, the bound \eqref{bound} is not meant to be optimal for any rank tensor models. Therefore, the above procedure might be
very well  improved in order to  investigate the finiteness or not of open amplitudes in any rank $d\geq  3$ theory. 

\begin{figure}[h]
 \centering
     \begin{minipage}{.5\textwidth}
\includegraphics[angle=0, width=6cm, height=2.5cm]{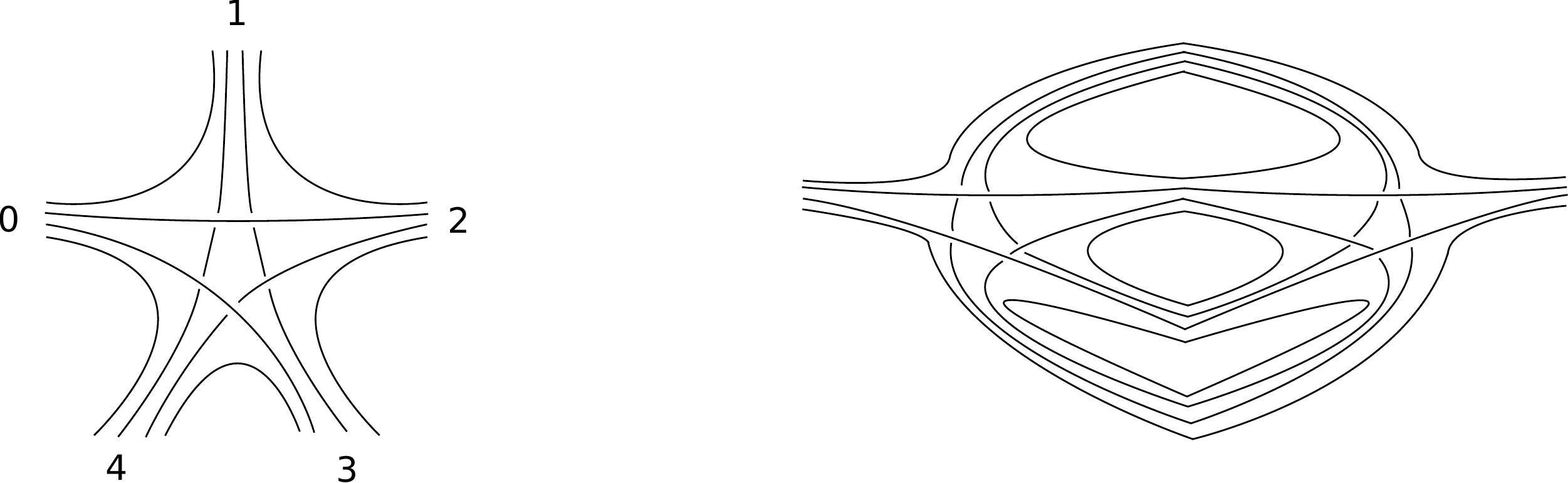} 
\caption{ {\small  Vertex representing a 4-simplex (left) and 2point-function of the Ooguri model (right). }} 
\label{fig:2pt}
\end{minipage}
\end{figure}

We can now illustrate the formula \eqref{bound} by considering the 
Abelian version of the rank $d=4$ Ooguri model, 
defined over $SO(4)$ relevant for 4 dimensional gravity \cite{Boul}.
The Ooguri propagator is a line with four strands and vertex
pattern follows the identification of triangles between 
5 tetrahedra in a 4-simplex (see Figure \ref{fig:2pt}).  
This model expands in terms of four dimensional simplicial complexes.  
Corresponding to an Abelian model version, we discuss the rank $d=4$ tensor model with group 
$G = U(1)^{D=6}$ and equipped with gauge invariant colored fields. 
Using \eqref{bound}, one computes  a bound on the 
two-point function amplitude corresponding to  $\cG_2$ (see Figure \ref{fig:2pt}) as 
(noting that, for $\cG_2$, one has  $F_{\inter,k\geq 4} =0$,  $\frac{1}{2} F_{\ext,1}=2$, $F_{\ext, k \neq  1}=0$, $D_{k \neq 4}=0$, $D_4=1$):
\bea
\omega^{\col}_{4}(T_2) \leq 10 
\label{t2}
\eea
which coincides with the initial superficial degree of divergence
$\omega^{\col}_{4}(\cG_2) = -2(4) + 6 (6-3)=10$. 

\section{Conclusion}
\label{ccl}

The Boulatov model \cite{Boul} generates three dimensional  simplicial complexes. It also provides  a topological theory with  
divergent amplitudes. At the perturbative
level, in the case of a trivial dynamics (usual model without Laplacian but just a mass-like term), 
there is, at this point, no way to provide a complete understanding of divergent amplitudes. 
By adding a Laplacian which proves to be naturally generated by quantum corrections in the model \cite{Geloun:2011cy} plus additional features like colors and
linearizing the group, it turns out that the simplicial complexes with boundaries have all now a convergent amplitude. 
This finiteness only happens in the rank 3 case, higher rank situations which we have studied as well are more involved and may still have divergent open graphs. 

Thus, in the rank 3 case, introducing a Laplacian dynamics can be interpreted as a UV-regulator for the theory
without this term\footnote{The author would like to thank Laurent Freidel for having pointed out this feature.}. 
We are in presence of two regimes. 
On one hand, we have a regime without Laplacian with uncontrollable divergences (leading to a non-renormalizable model) 
but which neatly describes a topological theory. On the other hand, by incorporating seemingly 
important quantum corrections in the form of a Laplacian in the initial action, the model becomes ``finite'' (in the sense discussed above)
and the topological property of the theory gets broken. 
The ensuing question is: ``Is there a field theory meaningful way to describe this new type of UV-regulator or does it make a 
sense to renormalize the theory using such a UV cut-off ?''. 
This unexpected and intriguing feature of that 3 dimensional gravity model must be given a proper 
field theory sense and fully deserves to be investigated elsewhere.

\section{Aknowledgments}
Insightful discussions with Laurent Freidel and Vincent Rivasseau are gratefully acknowledged.
 This research was supported in part by Perimeter Institute for Theoretical Physics. 
Research at Perimeter Institute is supported by the Government of Canada through 
Industry Canada and by the Province of Ontario through the Ministry of Research and Innovation.

\end{document}